\documentclass[12pt,preprint]{emulateapj}
\usepackage{psfig} 

\slugcomment{In press at ApJ.}

\shorttitle{SEDs of SMGs: a new {\it Spitzer} population?} 
\shortauthors{Blain et al.}

\begin{document}


\title{Accurate spectral energy distributions and selection effects for high-redshift dusty 
galaxies: a new\\ 
hot population to discover with the {\it Spitzer Space Telescope}?}   
\author{A.\,W. Blain and S.\,C. Chapman}
\affil{Caltech, Pasadena, CA 91125}  
\author{Ian Smail}
\affil{Institute for Computational Cosmology, University of Durham, Durham, UK} 
\and
\author{Rob Ivison\altaffilmark{1}}    
\affil{UK-ATC, Royal Observatory, Blackford Hill, Edinburgh, UK}
\altaffiltext{1}{Institute for Astronomy, University of Edinburgh, Blackford Hill, 
Edinburgh, UK}



\begin{abstract}
The spectral energy distributions (SEDs) of dust-enshrouded galaxies with 
powerful restframe far-infrared(IR) emission have been constrained 
by a range of ground-based and space-borne surveys. The 
{\it IRAS} catalog provides a 
reasonably complete picture of the dust emission from nearby galaxies 
(redshift $z \simeq 0.1$) that 
are typically less luminous 
than about $10^{12}\,L_\odot$. However, at higher redshifts, the 
observational 
coverage from all existing far-IR and submillimeter surveys is much 
less complete. Here we investigate the SEDs of  
a new sample of high-redshift 
submillimeter-selected galaxies (SMGs), for which 
redshifts are known, allowing us to estimate   
reliable luminosities and 
characteristic dust temperatures. We demonstrate that  
a wide range of SEDs is present in the population, and that 
a substantial number of luminous dusty galaxies with hotter 
dust temperatures could exist at similar redshifts  
($z \simeq 2-3$), but remain undetected in existing submillimeter 
surveys. These hotter galaxies could be responsible for 
about a third of the extragalactic IR background radiation at a 
wavelength of about 100\,$\mu$m. The brightest 
of these galaxies would have far-IR luminosities of the order 
of $10^{13}\,L_\odot$
and dust temperatures of the order of 60\,K. Galaxies  
up to an order of 
magnitude less luminous with similar SEDs will be easy to detect and 
identify in the deepest {\it Spitzer Space Telescope} 
observations of extragalactic fields at 24\,$\mu$m.  
\end{abstract}




\section{Introduction}

Deep submillimeter-wave surveys for distant galaxies offer direct access 
to high redshifts, because their sensitivity is almost independent of
galaxy luminosity for a fixed spectral energy distribution (SED) over the 
wide redshift range $1<z<5$ (Smail et al. 2002). The positional 
accuracy of these surveys is only $\simeq 8$\,arcsec, and optical 
counterparts are typically faint. Hence 
it has been difficult to identify and measure the redshifts of 
a substantial fraction of these submillimeter-selected galaxies (SMGs).  

These difficulties have now been overcome with 
rest-UV spectroscopy of SMGs 
located in very deep ($\simeq 10\,\mu$Jy) 
radio images by Chapman et al. (2003b; 2004a). Redshifts have been 
found for about half of the SMG population using  
the Keck I telescope's blue-sensitive multiobject spectrograph LRIS-B.
The resulting 
sample of 73 spectroscopic redshifts is very valuable, allowing a fairly 
accurate restframe SED to be determined for each galaxy, under 
reasonable assumptions about their radio--far-IR properties. 
The SED model is most strongly affected by the
value of the effective dust temperature of the galaxies $T_{\rm d}$, which can 
be treated as a proxy for the 
peak restframe wavelength of the SED. Just two parameters,  
$T_{\rm d}$ and the far-IR luminosity 
$L$\footnote{
Defined as 
the luminosity between wavelengths of 3\,mm and 10\,$\mu$m, and typically 
close to the bolometric luminosity.} 
allow the parameter space of SEDs that describe 
observational data 
at wavelengths longer than about 20\,$\mu$m to be spanned reasonably 
well (Dale et al.\ 2001; Blain, Barnard \& Chapman 2003; hereafter BBC03). 
Two other parameters are required to describe the detailed shape of the 
SED. The Rayleigh--Jeans spectral slope of dust emission is fixed by the  
emissivity index 
of dust $\beta$ 
($f_\nu \propto \nu^{2+\beta}$). The value of $\beta$ is degenerate with 
temperature, in the sense that models in which 
$T_{\rm d}\beta$ is approximately constant provide similarly accurate 
descriptions of the SED (see Fig.\,2 of BBC03). 
This effect could be important if $T_{\rm d}$ 
were treated as a real physical temperature, but when $T_{\rm d}$ is used 
as the key descriptive 
parameter in the SED the effect of this degeneracy between the fitted 
values of $T_{\rm d}$ and $\beta$ is not important. The slope of the 
mid-IR spectrum $f_\nu \propto \nu^\alpha$ is set by a second 
parameter, which has little effect on the value of $L$ at fixed $T_{\rm d}$. 

Spectroscopic redshifts are especially important for understanding 
the SEDs of SMGs as
there is a very significant, pernicious  
degeneracy between the inferred dust temperature $T_{\rm d}$ and 
redshift $z$, which only allows accurate measurement of the 
quantity $T_{\rm d}/(1+z)$ (Blain 1999; BBC03). 
The large Chapman et al. sample of secure spectroscopic redshifts 
for SMGs provides a way to break
this degeneracy, and thus unambiguously reveals the range of SED properties of the 
high-redshift population of SMGs. This is the 
population of galaxies for which radio--submillimeter--far-IR 
photometric redshifts are often sought (Carilli \& Yun 1998; 
Hughes et al.\ 2002; Yun \& Carilli 2002; Aretxaga 
et al. 2003; Wiklind 2003), and we  
use this new, accurate SED information, to address 
the precision of these radio--far-IR photometric redshift techniques.
We then compare the SEDs of low- and high-redshift samples of  
luminous dusty galaxies to 
highlight a region 
of the $L$--$T_{\rm d}$ parameter space where  
unrecognized populations could still lurk, and discuss ways to 
find them. 

Throughout this paper we assume a 
cosmology with $\Omega_0=0.3$, $\Omega_\Lambda=0.7$ and
$H_0=100h$\,km\,s$^{-1}$\,Mpc$^{-1}$ with $h=0.65$ to 
remain consistent with earlier work. In the {\it WMAP} 
cosmology with 
$\Omega_0=0.27$, $\Omega_\Lambda=0.73$ and
$h=0.71$ the 
quoted luminosities 
would be reduced by a factor of about 15\% at $z \simeq 2.5$.

\section{Luminosities and temperatures of dust-enshrouded galaxies} 

In Fig.\,1 we show values of inferred dust temperature $T_{\rm d}$ 
and bolometric 
far-IR luminosity $L$ 
for several samples of dust-enshrouded galaxies. These 
include well-studied low-redshift {\it IRAS} galaxies with submillimeter 
measurements (Dunne et al.\ 2000; Dunne \& Eales 2001), moderate-redshift  
galaxies selected jointly from the 
{\it IRAS} and the VLA-FIRST survey 
(Stanford et al. 2000) and high-redshift galaxies selected using both 
{\it Infrared Space Observatory (ISO)} 
at 15\,$\mu$m and radio images (Garrett et al. 2001).
Several examples of serendipitously selected 
high-redshift 
SMGs with known redshifts are also shown, including some 
gravitationally lensed examples. The values of $T_{\rm d}$ and $L$ 
inferred from these samples and their uncertainties 
were discussed 
by BBC03, and have now been updated for Fig.\,1. 
We now also plot the inferred $T_{\rm d}$ and $L$ values derived for the  
Chapman et al. (2003b, 2004a) 
sample of 73 SMGs with spectroscopic redshifts. 

Following 
BBC03, values of $T_{\rm d}$ and $L$
are derived assuming a modified blackbody 
SED with a single dust 
temperature, and a dust emissivity index $\beta=1.5$, 
which is extrapolated with a continuous gradient 
to a power-law $f_\nu \propto \nu^\alpha$ 
at frequencies above 
which the gradient of 
the blackbody spectrum is steeper than this power law. The 
value of $\alpha = -1.7$ is used here, because it provides a good 
description of the shape of 15-$\mu$m source counts based on the 
form of evolution inferred for the SMGs (Blain et al.\ 2002). More 
information about the range of values of this parameter awaits 
observations with the {\it Spitzer Space 
Telescope}.\footnote{http://ssc.spitzer.caltech.edu} 
Arp\,220 has amongst the steepest indices  
known with $\alpha = -2.9$ (BBC03). 

All the galaxy samples shown in Fig.\,1 have been fitted using the same 
function, so the derived luminosities and temperatures for each 
sample should be directly comparable. This three-parameter 
($T_{\rm d}i$, $\alpha$, $\beta$) model is 
the simplest that can 
describe the data adequately, and $T_{\rm d}$ is the most important 
parameter (BBC03). To emphasize, the 
value of $T_{\rm d}$ does not necessarily reflect the 
true physical temperature of most of the dust in the galaxy, but does 
provide a good description of the general properties of the SED.   

When fitting the SEDs, we 
assume that the low-redshift radio--far-IR flux density correlation 
(Condon 1992), resulting from the fact that high-mass stars are 
the common power source  
for both synchrotron radio emission via supernovae shock acceleration, 
and far-IR dust emission via absorbed blue/UV stellar light, applies 
to the Chapman et al. (2003b, 2004a) galaxies. Under this assumption, we 
can determine a temperature $T_{\rm d}$ to describe the SED from just two 
data points at 1.4\,GHz and 850\,$\mu$m. If there are systematic 
changes in the radio--far-IR flux density ratio or its scatter with redshift, 
then these values will be in error. So far there is no evidence 
for evolution in the form of the far-IR--radio correlation 
(Garrett et al.\ 2001).   
Information to test the inferred SEDs could be provided by difficult-to-obtain 
shorter submillimeter flux density measurements in the 350- and 450-$\mu$m 
atmospheric windows, 
or from space-borne 
far-IR data. So far, archival 450-$\mu$m SCUBA measurements at the 
James CLerk Maxwell Telescope 
(Smail et al.\ 2002) and new 350-$\mu$m 
measurements using the Caltech Submillimeter Observatory (Mauna Kea)'s 
SHARC-2 camera by Kovacs et al. (2004) for about 
10 SMGs 
show no major discrepancy as compared with the radio-submillimeter result: the 
dust temperatures inferred from the measured submillimeter color between 350 
and 850\,$\mu$m are consistent with those derived using the far-IR--radio 
correlation (Fig.\,1)  
to within the errors.  

The presence of an active galactic nucleus (AGN) might be 
expected to boost the radio flux density, leading to an overestimate 
of $T_{\rm d}$ (Yun, Reddy \& Condon 2001). This is clearly the case 
in only 3 out of 73 galaxies in this sample, for which very high dust 
temperatures and luminosities are inferred. Where X-ray observations are available for 
the Chapman et al. sample, the results are inconsistent with a 
large fraction of Compton-thin AGNs (Alexander et al.\ 2004).   

Note that, as discussed by BBC03, the addition of more
continuum data points 
reduces the uncertainties in the fits, but does not break the basic 
redshift--temperature degeneracy for SED fits. 

The size of the uncertainties on the estimated parameters $L$
and $T_{\rm d}$ for the Chapman et al. sources are 
highlighted in Fig.\,2 along with their redshifts: 
the uncertainties on the estimates for the other samples 
were discussed by BBC03. 
The individual uncertainty on 
any determination is comparable to the extent of the $L$--$T_{\rm d}$
parameter space spanned by the Chapman et al. sample. 
We also highlight the 
selection criteria for the SMGs in Fig.\,2. 
The error bars to the $T_{\rm d}$ are assigned by fitting 
SED templates to 
the colors of 
observed SMGs, using a maximum-likelihood estimator combined with 
broad priors, and without imposing any condition on luminosity. 
The error bars in luminosity are derived from the range of  
luminosity values for which $\chi^2 < 4$ for the two data 
points as the temperature moves away from the best-fitting value.   
Examples of correlations
between fitted values of $L$ and 
$T_{\rm d}$ are shown by a joint simple $\chi^2$ fit 
to the flux density data for two typical SMGs 
in Fig.\,3. 

\section{Selection effects} 

The different galaxy samples occupy somewhat different regions of the 
$L$-$T_{\rm d}$ plane shown in Fig.\,1. Part of this separation is 
due to their distinct 
selection criteria. The low-redshift {\it IRAS} 
galaxies occupy a relatively well-defined strip, whether their SEDs are  
determined from either 
far-IR--submillimeter colors (open squares; Dunne et al.\ 2000) 
or from radio--far-IR template fitting (solid lines; Chapman et al.\ 2003a).
The more distant Stanford et al.\ (2000) sample includes galaxies with 
greater luminosities, and with 
typically higher and more
scattered temperatures. These two samples match well 
at intermediate luminosities, $L \simeq 10^{12}\,L_\odot$, for which 
typical temperatures of about 40\,K are determined.  
It is likely that both of these
samples are biased against galaxies with cooler dust temperatures, 
because their inclusion requires a detection at the relatively short observed 
wavelength of 60\,$\mu$m. This 
corresponds to an even 
shorter restframe wavelength of 
about 45\,$\mu$m for the typical galaxy in the Stanford et al. sample.
A few faint galaxies selected jointly at radio and mid-IR wavelengths 
are found to have SEDs consistent with lower temperatures
(Garrett et al. 2001; 
see also 
Chapman et al.\ 2002).
The new, large and more 
accurately defined Chapman et al.\ (2004a) SMG sample (roughly 
detected at a restframe wavelength of 250\,$\mu$m)
also is systematically cooler, with a locus of points that 
lie at significantly lower temperatures than 
the trend of the {\it IRAS}-selected samples in the $L$--$T_{\rm d}$ 
plane shown in Figs\,1 \& 2. 

The track of the $L$--$T_{\rm d}$ distribution of 
SMG points in Fig.\,2 appears to 
be relatively tight, but this does not necessarily 
reflect a tight intrinsic scatter in the properties of the 
SMGs. Rather, 
the shape of this distribution is  
determined mainly by the 
condition of a detectable submillimeter-wave flux density, as shown by the 
dashed curves that trace the limits to detection for a 
5-mJy 850-$\mu$m source in the Chapman et al. (2004a) SMG sample at a
variety of redshifts in Fig.\,2. Thus, it is 
impossible to select SMGs in the high-$T_{\rm d}$--low-$L$ 
region shown in the figures at 
any redshift in existing submillimeter-wave surveys. 
In a deeper survey, as will be possible with the Atacama 
Large Millimeter Array (ALMA; Wootten et al.\ 2001) the 
region of the parameter space that can be probed will expand, and 
many high-redshift galaxies will then 
probably be found in this region. 
The nature of high-redshift galaxies means that there are 
many possible explanations 
for why they avoid the low-luminosity--high-temperature 
region of the $L$--$T_{\rm d}$ parameter 
space. They could have systematically different dust grain properties and 
spatial distributions of dust 
as compared with their more luminous companions, or they could 
be dusty galaxies at the start or end of a burst of  
enhanced star-formation/AGN activity, either brightening from or 
fading into 
the distribution of much more normal quiescent galaxies. Note that regular 
star-forming Lyman-break galaxies (LBGs; Steidel et al.\ 2004) 
will be found somewhere 
in the $L$--$T_{\rm d}$ plane at high redshifts (Fig.\,2d). 
By detecting their 
lower   
far-IR luminosities using ALMA it should be possible to understand the 
currently ambiguous relationship between the LBGs and SMGs in more detail. 

The corresponding  
conditions for radio detection at 1.4-GHz -- that a galaxy must exceed a 
flux density of about 
30\,$\mu$Jy is also shown in Fig.\,2. 
The shape of the radio selection function is much 
less sensitive to temperature, depending much more strongly on 
luminosity, with  
a more intuitive redshift dependence. 

The absence of points at high-luminosities in the $L$--$T_{\rm d}$ 
diagram reflects the 
steep high-luminosity cutoff of the luminosity function of 
dusty galaxies -- there are few very luminous objects to find.  
Note that the observed strong luminosity evolution of dusty 
galaxies by a factor $(1+z)^\gamma$ with $\gamma \simeq 4$ for $z<2$ 
(Blain et al.\ 1999),  
is implicit in the results shown for different redshift ranges in 
Fig.\,2. Very 
luminous galaxies are not represented in large numbers in the 
Stanford et al. sample at $z \sim 0.5$, yet are abundant in the SMG 
sample at $z \simeq 2.5$. 

Chapman et al.\ (2004a) found wide discrepancies while 
comparing the scatter in the radio--submillimeter flux ratios for SMGs with 
redshifts to the ratio expected assuming 
the updated Carilli--Yun 1.4-GHz--850-$\mu$m redshift estimator 
(Yun \& Carilli 2002). 
The range of dust temperatures for SMGs shown 
in Fig.\,2 is spread over a factor of at least 
1.4 for 
SMGs with $L \simeq 10^{13}\,L_\odot$. This implies  
a radio--submillimeter--far-IR photometric 
redshift accuracy no better than $\Delta z \simeq 1$ 
based on the 
temperature--redshift degeneracy (BBC03) for the typical 
SMG spectroscopic redshift $z \sim 2.5$:
this is in marked contrast to recent claims of much more accurate 
photometric redshifts by 
Aretxaga et al.\ (2003).  
The presence of 
hotter dusty galaxies at slightly lower luminosities in the Stanford et al. (2000)  
sample (Fig.\,1) hints that photometric redshift errors 
could be even larger if SMGs were 
selected down to deeper submillimeter flux density limits. 
Photometric redshift techniques at shorter near- and mid-IR wavelengths offer 
better hope of progress, either using redshifted polycyclic 
aromatic hydrocarbon (PAH)/silicate dust grain 
absorption and emission features, 
or stellar spectral features (Simpson \& Eisenhardt 1999; Sawicki 2002; 
Efstathiou \& Rowan-Robinson 2003).  
Regular multi-band near-IR, optical, UV photometric redshift techniques 
also hold promise, 
as many SMGs have the expected optical colors for their
spectroscopic redshift (Chapman et al.\ 2004a), despite having far-IR
luminosities that are almost impossible to predict from 
the optical data. Nevertheless, accurate positions, perhaps from radio 
or ALMA observations, are still necessary in order to confirm that a 
particular faint 
galaxy with a photometric redshift really is the counterpart to a 
particular SMG. 

There is almost no 
overlap in physical properties between well-studied low-redshift {\it IRAS} 
galaxies and the existing catalogs of SMGs, as shown in Figs\,1 \& 4. 
These classes 
of galaxies occupy totally disjoint regions of the $L$--$T_{\rm d}$ 
parameter space, being  
separated by a factor of 30\% in temperature
at comparable luminosities, and by a factor of 10 in luminosity at 
comparable temperatures. In part this is a selection effect, 
imposed by the 
requirement of a 850-$\mu$m flux density in 
excess of 5\,mJy at low redshifts and/or luminosities, 
as shown in Fig.\,2. However, the lack of overlap in the 
figure could also be a real astrophysical effect, 
perhaps reflecting either a more diffuse distribution 
of dust and stars or a lower typical  
interstellar medium metallicity in the high-redshift SMGs as compared 
with the {\it IRAS} galaxies. Spatially-resolved 
imaging of the radio, submillimeter or far-IR 
emission from these galaxies will be necessary to reveal the true 
relationship between these populations, its causes and the 
relative astrophysics of the {\it IRAS} and SMG galaxy populations. 

We emphasize that spectroscopic redshifts for SMGs are essential to draw all 
these conclusions: without them reliable temperature and thus luminosity 
information 
cannot be derived.

\section{New populations} 

A significant fraction ($\simeq 50$\%) 
of the SMG population with 850-$\mu$m flux densities 
in excess of 5\,mJy are included in the Chapman et al. (2004a) redshift 
sample (Fig.\,2).
Unless there is a significant systematic shift in the SEDs of SMGs 
at fainter flux densities ($\simeq 1$\,mJy), then slightly fainter, 
less luminous SMGs 
will typically populate the region of Fig.\,1 that lies between 
the existing SMG samples and 
the Dunne et al. (2000) low-redshift {\it IRAS} 
galaxies, providing a smooth distribution 
in luminosity 
between the extremely luminous high-redshift galaxies and more ordinary 
local dusty galaxies. These fainter SMGs at moderate and 
high redshifts 
will be hard to
detect at submillimeter wavelengths prior to the commissioning of ALMA.

High-redshift dusty galaxies with fainter 850-$\mu$m flux densities 
may have luminosities similar to those of 
known SMGs, but higher temperatures. 
These would be an important class of objects, both very luminous
and hotter than any population detected so far (Fig.\,1). 
The {\it IRAS} survey 
is not deep enough to find such galaxies (Stanford et al.\ 2000), 
and small {\it ISO} samples do not appear to detect them in large 
numbers
(Garrett et al. 2001; Franceschini et al. 2003). 
The deepest {\it ISO} 15-$\mu$m images (Altieri et al.
1999; Metcalfe et al. 2003) may include some examples, but  
redshifts, and supporting deep submillimeter and radio data are not common for 
this population. 

With $T_{\rm d} \simeq 60$K and $L \simeq 10^{13}\,L_\odot$ such hot galaxies 
at $z \simeq 2$--3 could be detected in current deep radio surveys (Fig.\,2), 
but not in existing submillimeter surveys. 
A factor of 10 increase in submillimeter sensitivity 
is required to probe this region of the $L$--$T_{\rm d}$ diagram. 
Because confusion noise sets the 850-$\mu$m sensitivity limit  
at about 2\,mJy in field galaxy surveys using 
current telescopes 
(Blain et al.\ 2002), 
higher angular resolution 
or a large gravitational lensing 
magnification, not just longer integrations,  
would be necessary.

Optical spectra taken  
in parallel to Chapman et al.'s SMG redshift survey 
have provided redshifts for many optically faint 
galaxies that have weak radio detections but no
submillimeter emission (Chapman et al.\ 2004b). These galaxies' 
optical spectral 
properties are similar to those of high-redshift SMGs, without  
signs of powerful AGN, so it 
is plausible to suggest that their radio emission is produced 
by star formation. 
These optically faint galaxies are excellent candidates for a
hotter, high-redshift, dust-enshrouded galaxy population, which cannot 
be detected in submillimeter surveys owing to the galaxies' 
high dust temperatures.  
The surface 
densities of these galaxies are comparable to those of 
the SMGs, so a survey 
covering just a few tenths of a square degree should find a statistical 
sample of several tens. 

Such a population of galaxies would generate 
background radiation intensity  
in addition to that from the known SMGs and 
their fainter counterparts.  
The extra intensity that they contribute would mostly increase 
the background intensity at wavelengths shorter than about 200\,$\mu$m, 
where the background reported by 
Schlegel, Finkbeiner \& Davis (1998) is 
underpredicted both by models of the evolution of SMGs (Blain et al.\ 
1999, 2002), and by 
summing over the flux and  
surface densities reported
for SMGs and their fainter dusty counterparts. Possible sources of 200-$\mu$m 
background radiation are discussed by Blain \& Phillips (2002): a 
50\% increase in the intensity of background radiation above that 
accounted for by extrapolating the known SMG population is possible 
at about 100\,$\mu$m, and could be due to a hotter high-redshift 
population of dusty galaxies. 

The known selection effects for existing 
submillimeter and far-IR surveys, observational 
hints at 
a spectroscopically similar, but  
submillimeter-faint, population of high-redshift radio source, and the 
intensity of the 100--200-$\mu$m extragalactic background radiation,   
are all consistent with the existence of a proposed 
population of hot high-redshift dusty galaxies. 
To be certain that we have revealed this population 
it is necessary to probe deeper into the luminosity function of 
dusty galaxies at high redshifts, an 
opportunity that is available most easily  
at shorter mid-IR wavelengths. 

\section{{\it Spitzer} probes of new populations} 

The recently launched {\it 
Spitzer}
offers excellent sensitivity and an unprecedentedly 
large field of view at mid and far-IR wavelengths. 
While it is still not absolutely clear 
how deep {\it Spitzer} can integrate 
before being limited by confusion 
noise, it will certainly produce much better images of 
the far-IR sky 
than any previous mission. Fig.\,4 shows the region of the $L$--$T_{\rm d}$ plane that 
{\it Spitzer} will probe to the estimated confusion limits (Blain et al.\
2002) at various wavelengths. 
Deep {\it Spitzer} observations can thus investigate 
the whole of the unprobed region of the 
$L$--$T_{\rm d}$ plane near 60\,K and $10^{13}\,L_\odot$ at 24-$\mu$m. 
In the unlikely event that a source confusion limit brighter than 
0.1\,mJy prevents 
{\it Spitzer} from reaching this depth, then 
the larger 2.5-m aperture SOFIA\footnote{http://sofia.arc.nasa.gov} 
airborne observatory 
will certainly be able to probe this region of the $L$--$T_{\rm d}$ plane
in very long exposures. 

The 24-$\mu$m flux density expected for a 
distant galaxy as a function of $T_{\rm d}$ and $L$ depends most 
strongly on the value of the $\alpha$ parameter that describes 
the mid-IR 
SED. However, increasing or decreasing $\alpha$ by 0.5 leads to less 
than a factor of two change in the predicted far-IR luminosity at 
24-$\mu$m expected from 
a galaxy at redshift $z=2.5$ for $T_{\rm d}>40$\,K. 
If an extreme value of $\alpha = -2.9$ 
(representative of the steepest known mid-IR SED in Arp\,220) is 
assumed, then 
only galaxy luminosities that are a factor of 5 times higher than the 
limits shown in Fig.\,4 can be probed 
using a 24-$\mu$m observation to a certain flux density limit.  
Hence, even for the most extreme SED, 
a {\it Spitzer} 24-$\mu$m survey can still access the unexplored 
region of the $L$--$T_{\rm d}$
parameter space highlighted in Fig.\,4 for galaxies 
at $z \simeq 2.5$.   

At wavelengths longer than about 100\,$\mu$m, confusion noise from faint
undetected galaxies
makes it difficult to probe the unexplored region at
high redshifts unless the telescope used has
an aperture greater than of order 3-m. Hence, to sample 
the full range of known 
high-redshift galaxy SEDs near their restframe emission peak 
will require a next-generation 10-m-class far-IR 
telescope, such as the NASA {\it SAFIR} vision 
mission.\footnote{http://safir.gsfc.nasa.gov}   

Its sensitivity to less-luminous and hotter 
galaxies at lower redshifts 
will also ensure that {\it Spitzer} can be used 
to reveal whether the absence of 
low-luminosity, hot, relatively low-redshift 
galaxies in  
Fig.\,4 reflects a real upper limit to the luminosity-weighted 
dust temperature in  
galaxies, or whether it is 
a selection effect brought about by insufficient sensitivity in 
existing surveys. Note that because the line defining the 24-$\mu$m 
selection condition is predicted to be orthogonal to 
the track of the existing 
data points through Fig.\,4, {\it Spitzer} observations will be sensitive 
to any high-redshift galaxies with luminosities in the 
range $10^{11}$--$10^{12}\,L_\odot$, and temperatures $T_{\rm d}>50$\,K. 
Hence, 
they could readily test whether the upper temperature limit to the 
SEDs in existing samples is a 
real astrophysical result or a selection effect. 

A further benefit of the great sensitivity of 
{\it Spitzer} is its ability to detect any luminous 
galaxies 
that lie on the cold side 
of the SMG track in 
Fig.\,4 
(e.g. Chapman et al. 2002).
The details of the detectability of cooler galaxies at the 
short rest wavelength of 24\,$\mu$m are uncertain, and will 
remain so until the  
release of the first {\it Spitzer} data in mid-2004. 
Although such cold luminous galaxies should be detectable in existing submillimeter 
surveys, the much larger sky area that should be covered by {\it Spitzer} -- 
of the order of 
100\,deg$^2$ as compared with the 0.5--1\,deg$^2$ covered 
by all existing deep mm and submillimeter-wave surveys -- will 
yield a larger number of potential examples. 
Deep {\it Spitzer} images of SMG fields will also provide a direct 
test of the SEDs assumed for the SMGs based on the far-IR--radio 
correlation 
when placing the SMGs on the 
$L$--$T_{\rm d}$ plane in 
Fig.\,1, and thus will test the evolution of this correlation. 
Note that to identify the most interesting sources among 
the huge crop of {\it Spitzer} detections, deep radio  
and submillimeter data will still be very valuable,  
to provide accurate positions and SED/temperature limits  
respectively. 

If {\it Spitzer} detects faint high-redshift $\mu$Jy radio 
galaxies and confirms that they lie in the unexplored 
region of the $L$--$T_{\rm d}$ plane (Fig.\,4), 
then we will have a full measure 
of the scatter 
in the SEDs/temperatures of high-redshift luminous dusty galaxies. 
If not, then we
will have found a new 
population of distant radio-selected galaxies without 
obvious AGN signatures (Chapman et al.\ 2004b; Alexander 
et al.\ 2004). These 
galaxies would have to lie on the radio-loud side of the low-redshift 
radio--far-IR correlation. 
This new 
class of galaxies could be either X-ray and 
optically faint, dust-enshrouded, Compton-thick AGNs, or distant 
star-forming galaxies with 
anomalously powerful radio emission, which are found only 
at the 1\% level 
in low-redshift samples (Yun et al.\ 2001).  

\begin{figure*}[ht]
\begin{center}
\begin{tabular}{c}
\psfig{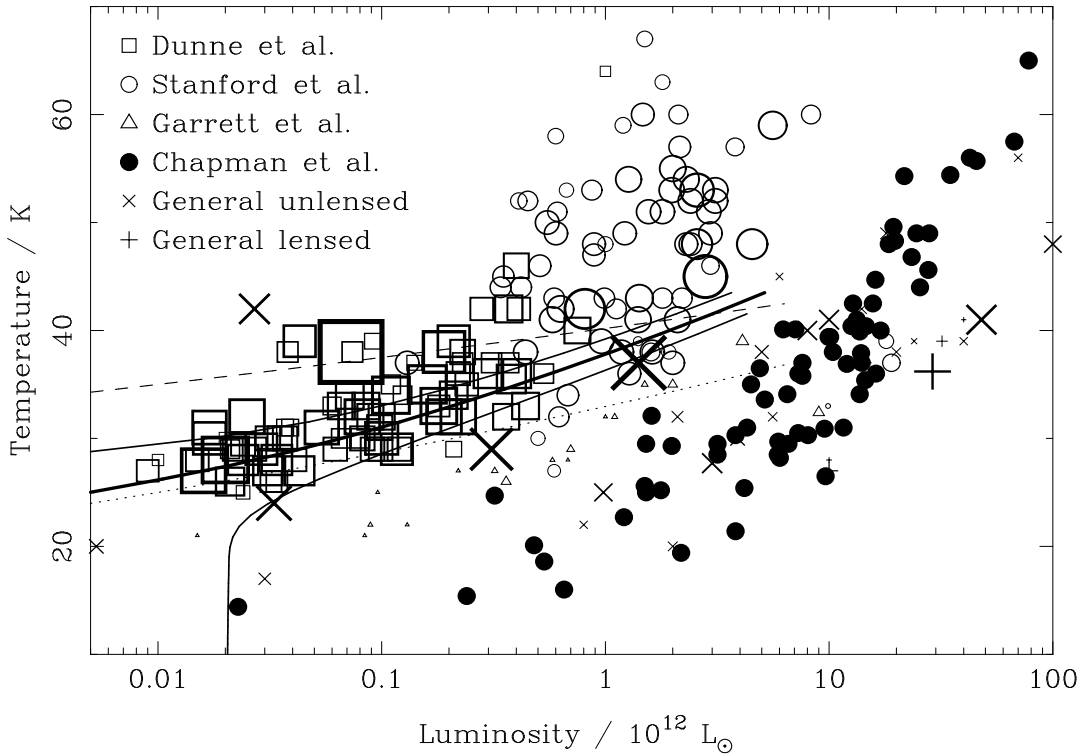} \\
\end{tabular}
\end{center}
\smallskip
\caption{Range of luminosity $L$ and temperature $T_{\rm d}$ 
values determined for well-studied dusty galaxies at 
a variety of redshifts. Open squares, low-redshift {\it IRAS} galaxies  
(Dunne et al.\ 2000); open circles, moderate-redshift very luminous {\it IRAS} 
galaxies (Stanford et al.\ 2000); open triangles, high-redshift {\it ISO} 
galaxies (Garrett et al. 2001); filled circles, spectroscopically identified SMGs 
(Chapman et al. 2003b, 2004a). Crosses and plus signs show galaxies with 
well-determined 
SEDs from a wide variety of sources (including SCUBA and 
{\it ISO} surveys). Crosses represent 
a selection of gravitational lenses, while plus signs represent 
unlensed galaxies (for both see 
BBC03); and spectroscopically identified SMGs -- filled circles (Chapman 
et al.\ 2003, 2004a). 
The sample represented by diagonal crosses includes M82, the Milky Way, 
NGC\,958 and Arp\,220, which are shown by the largest crosses between 
$2 \times 10^{10}\,L_\odot$ and $1.6 \times 10^{12}\,L_\odot$ 
respectively. 
For the non-Chapman samples larger symbols reflect 
more accurate determinations: see BBC03 for a description of the errors for 
these samples. See Fig.\,2 for uncertainties in the estimates for 
the Chapman et al. sample, 
for which all symbols are the same size. 
The thick solid line 
shows the result derived by Chapman et al. (2003a) for the $L$--$T_{\rm d}$
values of low-redshift 
{\it IRAS} galaxies. The interquartile range for this sample 
is bracketed by the 
two thinner solid lines, the separation of which agrees well with the 
range of properties of the Dunne et al. sample.  
The dashed and dotted lines show the fits 
obtained to merging and quiescent low-redshift {\it IRAS} galaxy 
data respectively by Barnard (2002). 
\label{fig1}}
\end{figure*}

\begin{figure*}
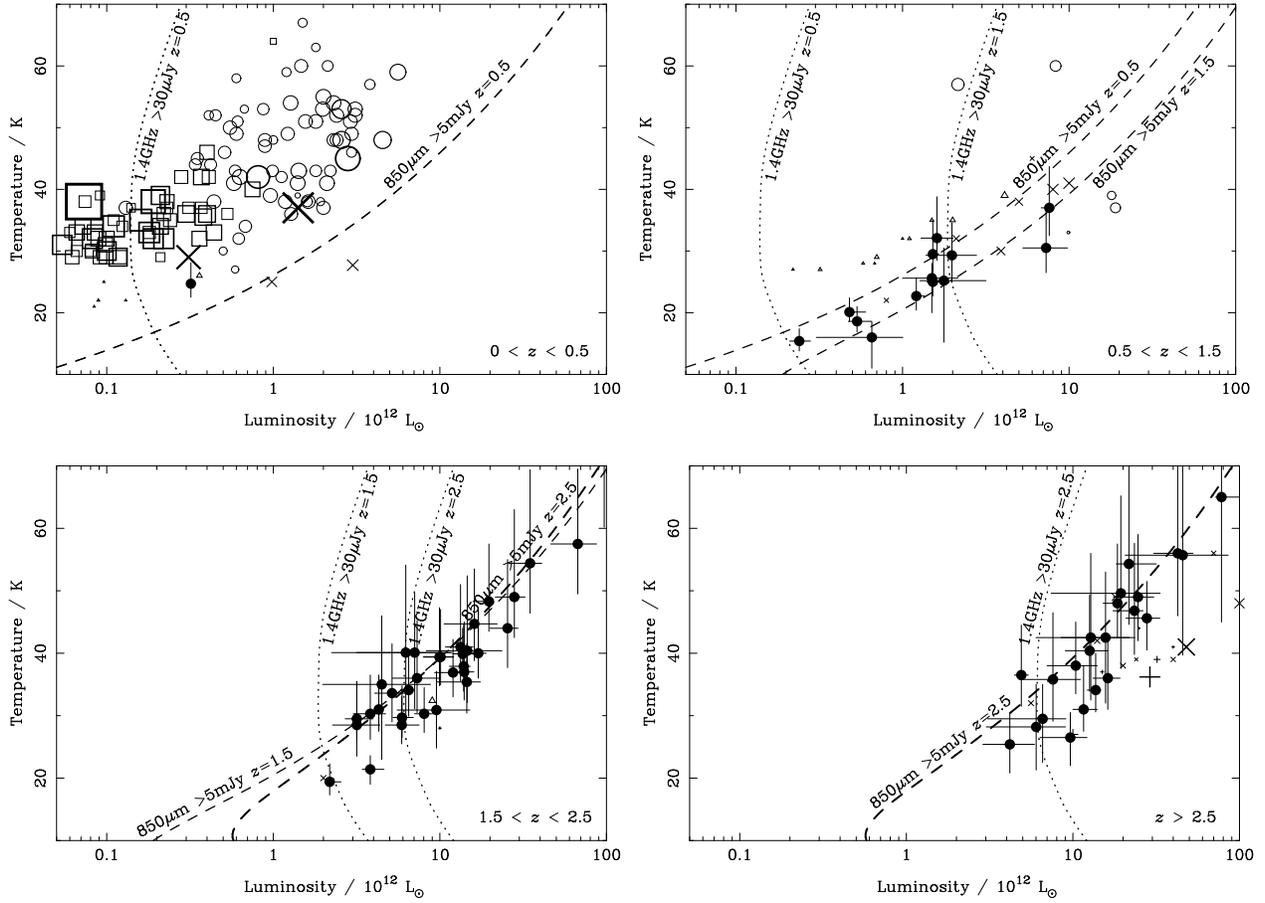

\begin{center}
\begin{tabular}{cc}
\psfig{figure=F2arev.eps,angle=-90,width=3.2in} \hspace{50pt}&
\psfig{figure=F2brev.eps,angle=-90,width=3.2in} \vspace{10pt} \\
\psfig{figure=F2crev.eps,angle=-90,width=3.2in} \hspace{50pt}&
\psfig{figure=F2drev.eps,angle=-90,width=3.2in} \\
\end{tabular}
\end{center}
\caption{SED parameters for the galaxies  
shown in Fig.\,1 divided by redshift range, and the errors 
on the parameters fitted to the Chapman et al. (2003b, 2004a) sample, symbols 
are as in Fig.\,1. The selection 
effects that limit the SED parameters for galaxies in this sample are 
represented by the lines at submillimeter and radio wavelengths for the 
labeled redshifts. Galaxies can be detected to the high-luminosity 
side of each line. 
There is a trend for the 
most luminous galaxies 
to be at the highest redshifts, reflecting the strong  
evolution of the luminosity function of the  
dust-enshrouded galaxy population. Nevertheless, 
note that broad ranges of luminosity, and more importantly temperature,
are sampled in each redshift interval regardless of the selection 
effects. 
\label{fig2}}
\end{figure*} 

\begin{figure*}
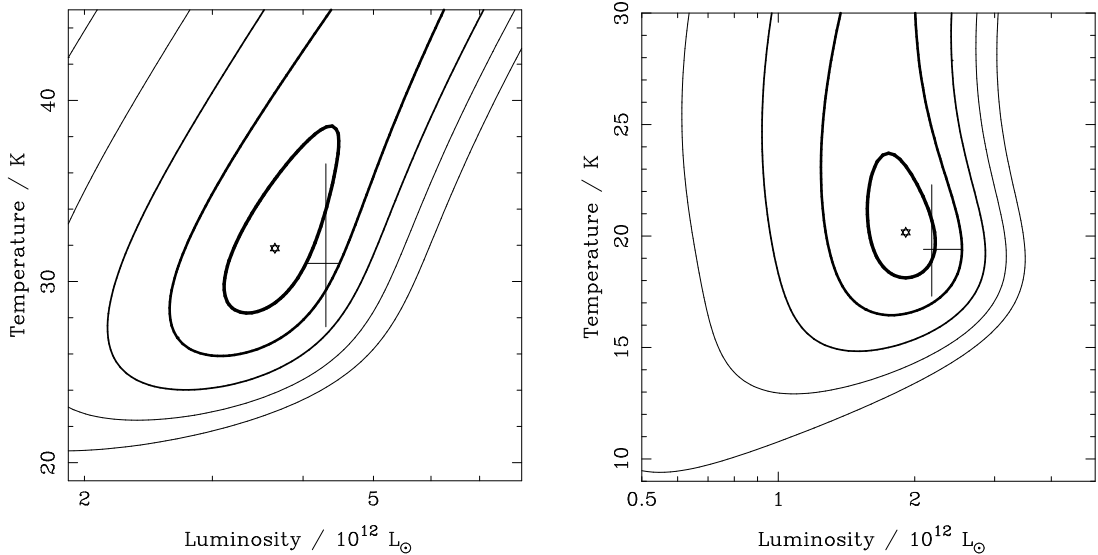

\begin{center}
\begin{tabular}{cc}
\psfig{figure=F3arev.eps,angle=-90,width=2.7in} \hspace{15pt} &
\psfig{figure=F3brev.eps,angle=-90,width=2.7in} \\
\end{tabular}
\end{center}
\caption{Examples of the uncertainties in the SED fit obtained for two 
typical radio-submillimeter 
galaxies in the HDF field. Left: SMMJ123707.21+621408.1, No. 242; right:
right is SMMJ123721.8+621035.3, No. 378) from 
Chapman et al.\ (2004a), at $z=2.48$ and $1.51$ respectively). 
The error bars match the results plotted in Fig.\,2, and are 
determined by making a full maximum-likelihood fit to the 
temperature $T_{\rm d}$
for an SED of arbitrary luminosity and then finding the range of 
best-fitting values of $L$ 
that have a value of $\chi^2<4$ at that value of $T_{\rm d}$. 
The contours show reduced 
$\chi^2$ values of 1, 4, 9, 16 and 25 derived directly for 
comparison. The   
$L$--$T_{\rm d}$ values are consistent, but slightly distinct, 
reflecting the different fitting procedures.  
The minimum $\chi^2$ value is denoted by the star. 
Note that the extent of the contours in the 
temperature direction can be reduced by adding any data from {\it 
Spitzer} mid- or far-IR observations. 
\label{fig3}}
\end{figure*}

\begin{figure*}
\begin{center}
\begin{tabular}{c}
\psfig{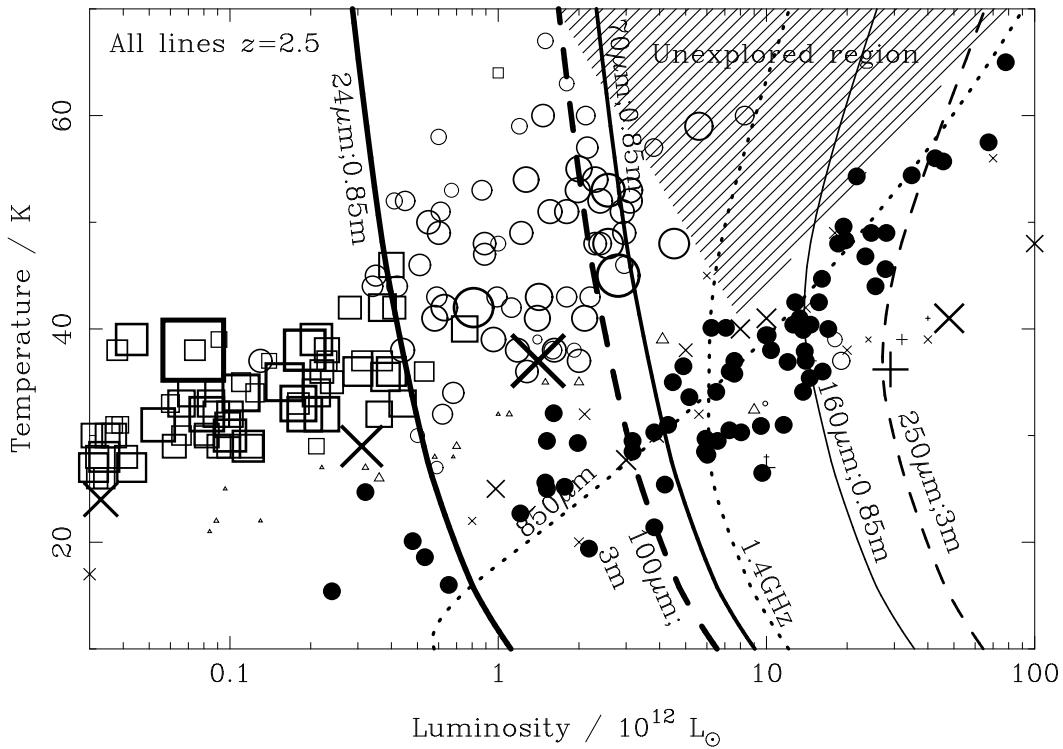}\\
\end{tabular}
\end{center}
\caption{Full range of $L$--$T_{\rm d}$ 
values determined for dusty galaxies 
(see Fig.\,1). The lines show the limits for selecting galaxies in 
surveys close to the 
confusion limit for a 0.85-m aperture telescope like {\it 
Spitzer} and a 3-m-class telescope like {\it Herschel} or SOFIA,
for a galaxy at redshift $z=2.5$, typical of the identified SMGs.
The solid lines show the limits of 
{\it Spitzer} at 24\,$\mu$m (thick lines; 0.1\,mJy), 70\,$\mu$m 
(medium lines; 5\,mJy) and 160\,$\mu$m (thin lines; 80\,mJy). 
The dashed lines represent the limits of a 3-m aperture 
telescope  
at 250\,$\mu$m (thin lines; 70\,mJy), and 100\,$\mu$m (thick lines; 1.5\,mJy). 
The selection limits in the 
existing Chapman et al. samples (Fig.\,2) 
are shown by the dotted 
lines. 
Galaxies to the high-$L$ side of all the lines can be detected. 
Reaching the unexplored region to find 
ultraluminous ($L \sim 10^{13}\,L_\odot$) galaxies at 
$T_{\rm d} \simeq 60$K
requires {\it Spitzer} observations 
at 24\,$\mu$m or an aperture larger than about 3\,m at 
observing wavelengths
longer than 100\,$\mu$m. 
To probe a significant part of the unexplored region of the diagram 
at submillimeter wavelengths would require 
a survey depth of order 0.1\,mJy at 850\,$\mu$m with ALMA.  
\label{fig4}}
\end{figure*}

\section{Conclusions}

The recent acquisition of a large sample of SMG redshifts  
provides the opportunity to constrain the SEDs of the most  
luminous ($> 10^{12}\,L_\odot$) high-redshift galaxies. 
It is now reasonably certain that the typical high-redshift SMGs selected 
at 850-$\mu$m are systematically 
cooler than local dusty galaxies
of comparable luminosity, and are thus without direct
low-redshift 
analogs. The scatter in their dust 
temperatures is about 50\%, sufficient to 
introduce a $\sim 30\%-40$\%  
uncertainty in {\emph any} photometric redshifts 
derived from 
far-IR, submillimeter, or radio 
photometry, no matter how accurate. 
The true scatter of the dust temperatures in high-redshift 
dusty galaxies certainly could 
be still greater, but hotter examples cannot be selected by 
existing submillimeter surveys. 
Faint radio galaxies that are undetected in the submillimeter, but whose 
optical properties are 
like those of the SMGs hint at the existence of a hotter population, 
with luminosities $L \simeq 10^{13}\,L_\odot$ and temperatures 
$T_{\rm d} \simeq 60$\,K. These galaxies would produce a 
luminosity density comparable to that of the 
existing populations of SMGs. 
The deepest 
forthcoming mid-IR {\it Spitzer} surveys should soon confirm whether this 
population exists, while providing a complete inventory of the range of 
SEDs present in high-redshift dusty galaxies in fields that have 
excellent supporting long-wavelength data. 
 
\acknowledgements

We thank Hilke Schlichting for some of the $L$--$T_{\rm d}$ results for the 
submillimeter galaxy population as part of her Caltech SURF program, and an 
anonymous referee for their helpful comments on the manuscript. A. W. B.  
is supported by NSF grant AST-0205937 and the Alfred P. Sloan Foundation. 
I. R. S. acknowledges support from 
the Royal Society.

\end{document}